# Building a Secure Software Supply Chain with GNU Guix

## Ludovic Courtès[a]

a    Inria, France

**Abstract**    The *software supply chain* is becoming a widespread analogy to designate the series of steps taken to go from source code published by developers to executables running on the users' computers. A security vulnerability in any of these steps puts users at risk, and evidence shows that attacks on the supply chain are becoming more common. The consequences of an attack on the software supply chain can be tragic in a society that relies on many interconnected software systems, and this has led research interest as well as governmental incentives for supply chain security to rise.

GNU Guix is a software deployment tool and software distribution that supports provenance tracking, reproducible builds, and reproducible software environments. Unlike many software distributions, it consists exclusively of source code: it provides a set of package definitions that describe how to build code from source. Together, these properties set it apart from many deployment tools that center on the distribution of binaries.

This paper focuses on one research question: how can Guix and similar systems allow users to securely update their software? Guix source code is distributed using the Git version control system; updating Guix-installed software packages means, first, updating the local copy of the Guix source code. Prior work on secure software updates focuses on systems very different from Guix—systems such as Debian, Fedora, or PyPI where updating consists in fetching metadata about the latest binary artifacts available—and is largely inapplicable in the context of Guix. By contrast, the main threats for Guix are attacks on its *source code repository*, which could lead users to run inauthentic code or to downgrade their system. Deployment tools that more closely resemble Guix, from Nix to Portage, either lack secure update mechanisms or suffer from shortcomings.

Our main contribution is a model and tool to authenticate new Git revisions. We further show how, building on Git semantics, we build protections against downgrade attacks and related threats. We explain implementation choices. This work has been deployed in production two years ago, giving us insight on its actual use at scale every day. The Git checkout authentication at its core is applicable beyond the specific use case of Guix, and we think it could benefit to developer teams that use Git.

As attacks on the software supply chain appear, security research is now looking at every link of the supply chain. Secure updates are one important aspect of the supply chain, but this paper also looks at the broader context: how Guix models and implements the supply chain, from upstream source code to binaries running on computers. While much recent work focuses on attestation—certifying each link of the supply chain—Guix takes a more radical approach: enabling independent *verification* of each step, building on reproducible builds, "bootstrappable" builds, and provenance tracking. The big picture shows how Guix can be used as the foundation of secure software supply chains.

## The Art, Science, and Engineering of Programming

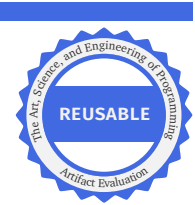

## 1 Introduction

Package managers and related software deployment tools are in a key position when it comes to securing the "software supply chain"—they take source code fresh from repositories and providing users with ready-to-use binaries. Between source code repositories and binaries users run, many things can go wrong: binaries can be compromised on their way to the user's machine [3], on the provider's servers, or possibly indirectly *via* toolchain compromission [43]. Every software installation and every upgrade can put users at risk. As the "last kilometer" of this supply chain, the way package managers distribute binaries and associated metadata led to security vulnerabilities that are now better understood and addressed [39]. But recent high-profile cases have reminded us that software supply chain attacks occurring *before* the distribution step are a very real threat [26, 35]. This led, for example, the US Government to call for work in this area in its Executive Order on cybersecurity, explicitly mentioning actions such as "using administratively separate build environments" and "employing automated tools (. . .) to maintain trusted source code supply chains" [1]. That there is room for improvement in current practices and tools is unquestioned.

GNU Guix is a set of software deployment tools and a standalone GNU/Linux distribution whose development started in 2012 [10]. It includes a package manager with a command-line interface similar to that of Debian's apt or Fedora's yum, allowing users to search for software packages, to install them, and to upgrade them. Unlike apt, yum, and many popular package managers, Guix builds upon the *functional deployment model* pioneered by Nix [17], a foundation for reproducible deployment, reproducible and verifiable builds, and provenance tracking. Guix is essentially a "source-based" deployment tool: the *model* is that of a system where every piece of software is built from source, and pre-built binaries are viewed as a mere optimization and not as a central aspect of its design.

This paper focuses on one research question: how can Guix and similar systems allow users to securely update their software? Guix source code is distributed using the Git version control system; updating Guix-installed software packages means, first, updating the local copy of the Guix source code. Prior work on secure software updates [25, 39] focuses on systems very different from Guix—systems such as Debian, Fedora, or PyPI where updating consists in fetching metadata about the latest binary artifacts available—and is largely inapplicable in the context of Guix. Deployment tools that more closely resemble Guix, from Nix to Portage and BSD Ports [7, 8, 14, 15, 17, 38], either lack secure update mechanisms or suffer from shortcomings. Git itself allows individual commits and tags to be authenticated but offers no way to authenticate a complete checkout—making sure *each* commit was made by an *authorized* party.

We describe the design and implementation of Guix's secure update mechanism. Section 2 gives background information necessary to understand the overall deployment model of Guix, showing how it supports *independent verification* of key links of the software supply chain. Section 3 presents our goals and threat model for the design of secure updates. Section 4 describes our design of a Git checkout authentication mechanism and Section 5 discusses trust establishment. Section 6 shows how we address downgrade attacks while Section 7 focuses on the related risk of distributing stale revisions. In Section 8 we provide key elements of the implementation and report on our experience. Last, Section 9 compares to related work and Section 10 concludes.

## 2 Background

Users of free operating systems such as GNU/Linux are familiar with *package managers* like Debian's `apt`, which allow them to install, upgrade, and remove software from a large collection of free software packages. GNU Guix[1] is such a tool, though it can be thought of more broadly as a toolbox for a software deployment with salient features and processes that improve security: a foundation for *reproducible builds*, and what we call *bootstrappable builds*.

### 2.1 A Deployment Toolbox

Guix provides a command-line interface similar to that of other package managers: `guix install python`, for instance, installs the Python interpreter, `guix pull` updates Guix itself and the set of available packages, and `guix upgrade` upgrades previously-installed packages to their latest available version. Package management is per-user rather than system-wide; it does not require system administrator privileges, nor does it require mutual trust among users.

Providing more than 20,000 software packages today, Guix is used as a general purpose day-to-day GNU/Linux distribution that provides the additional safety net of *transactional upgrades and rollbacks* for all software deployment operations. For example, if an upgrade has undesired effects, users can run `guix package --roll-back` to immediately restore packages as they were before the upgrade. Its ability to reproduce software environments, bit for bit, at different points in time and on different machines, makes it a tool of choice in support of reproducible computational experiments and software engineering [24].

Guix can be used on top of another system; the only requirement is that the system runs the Linux kernel—be it Android or a GNU/Linux distribution. Guix packages stand alone: they provide all the user-land software they need, down to the C library; this guarantees they behave the same on any system, as evidenced by more than twenty years of experience with the functional deployment model [10, 17].

There are other tools beyond the "package manager" interface. The `guix shell` command, for example, creates a one-off development environment containing the given packages. The `guix pack` command creates standalone *application bundles* or *container images* providing one or more software packages and all the packages they depend on at run time. The container images can be loaded by Docker, podman, and similar "container tools" to run the software on any other machine.

Last, Guix can be used as a standalone GNU/Linux distribution called Guix System. Its salient feature is that it lets users declare the *whole system configuration*—from user accounts, to services and installed packages—using a domain-specific language (DSL) embedded in Scheme, a functional programming language of the Lisp family [40]. The `guix system reconfigure` command changes the running system to match the user-provided configuration. This is an atomic operation and users can always roll back to an older "generation" of the system, should anything go wrong. The `guix system image` command can create system images in a variety of formats, including the QCOW2 format commonly-used for virtual machines (VMs) and emulators such as QEMU. `guix deploy` goes a step further and can deploy Guix System *on a set of machines*, be it over secure shell (SSH) connections or using the interfaces of a virtual private server (VPS) provider.

---

[1] `https://guix.gnu.org` (last accessed June 2022)

## 2.2 Reproducible Builds

At its core, Guix is a *functional* deployment tool that builds upon the ideas developed for the Nix package manager by Dolstra *et al.* [10, 17]. The term "functional" means that software build processes are considered as *pure functions*: given a set of inputs (compiler, libraries, build scripts, and so on), a package's build function is assumed to always produce the same result. Build results are stored in an immutable persistent data structure, the *store*, implemented as a single directory, `/gnu/store`. Each entry in `/gnu/store` has a file name composed of the hash of all the build inputs used to produce it, followed by a symbolic name. For example, `/gnu/store/yr9rk90jf...--gcc-10.3.0` identifies a specific build of GCC 10.3. A variant of GCC 10.3, for instance one using different build options or different dependencies, would get a different hash. Thus, each store file name uniquely identifies build results. This model is the foundation of *end-to-end provenance tracking*: Guix records and uniquely identifies the inputs leading to build results available in `/gnu/store`.

Guix, like Nix and unlike Debian or Fedora, is essentially a *source-based distribution*: Guix package definitions describe how to build packages from source. When running a command such as `guix install gcc`, Guix proceeds as if it were to build GCC from source. As an optimization, users can enable fetching pre-built binaries—called *substitutes* because they are substitutes for a local build. In that case, instead of building locally, Guix asks one or more servers for substitutes. In the example above, it would ask specifically for substitutes for `/gnu/store/yr9rk90jf...-gcc-10.3.0`, which unambiguously identifies the desired build output. Substitutes are cryptographically signed by the server and Guix rejects substitutes not signed by one of the keys the user authorized.

To maximize chances that build processes actually look like pure functions, they are spawned in isolated build environments—Linux *containers*—ensuring that only explicitly declared inputs are visible to the build process. This method, inherited from Nix [17], is illustrated in Figure 1: `guix` commands make remote procedure calls (RPCs) to a build daemon, which spawns build processes in isolated environments on their behalf and stores the build result in `/gnu/store`.

Build isolation, in turn, helps achieve bit-for-bit *reproducible builds*, which are critical from a security standpoint [26]. Reproducible builds enable users and developers to verify that a binary matches a given piece of source code: anyone can rebuild

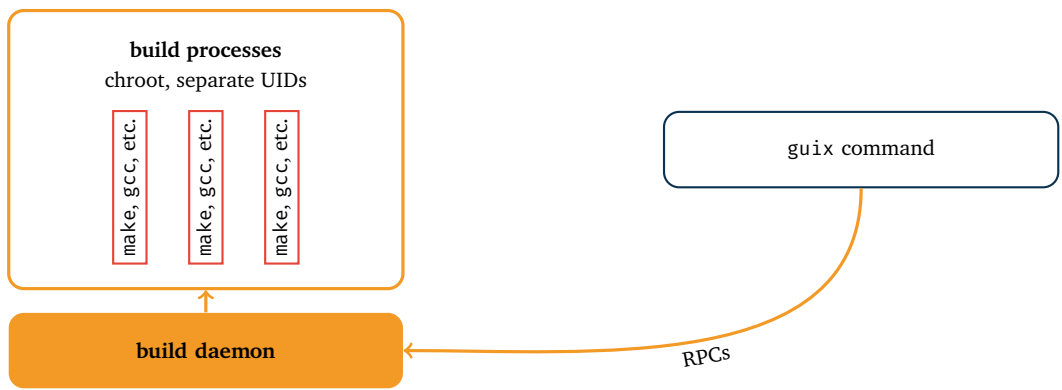

■ **Figure 1** The `guix` command makes remote procedure calls (RPCs) to a build daemon, which spawns hermetic builds on its behalf.

the package and ensure they obtain the same binary, bit for bit. The explicit and unambiguous mapping from source to binary that the functional deployment model provides makes verification clear and easy.

For example, the command `guix build --check hello` rebuilds the `hello` package locally and prints an error if the build result differs from that already available. Likewise, `guix challenge hello` compares binaries of the `hello` package available locally with those provided by one or several substitute servers. These two commands allow users and developers to find about binaries that might have been tampered with. For packagers, they are more commonly a way to find out about non-deterministic build processes—e.g., build processes that include timestamps or random seeds in their output, or that depend on hardware details.

## 2.3 Bootstrappable Builds

Are reproducible builds enough to guarantee that one can verify source-to-binary mappings? In his Turing Award acceptance speech, Ken Thompson described a scenario whereby a legitimate-looking build process would produce a malicious binary [43]. If that build process is reproducible, it just reproducibly builds a malicious binary. The attack Thompson described, often referred to as a "Trusting Trust attack", consists in targeting the compilation toolchain, typically by modifying the compiler such that it emits malicious code when it recognizes specific patterns of source code. This attack can be undetectable. What makes such attacks possible is that users and distributions rely on opaque binaries at some level to "bootstrap" the entire package dependency graph.

GNU/Linux systems are built around the C language. At the root of the package dependency graph, we have the GNU C Library (glibc), the GNU Compiler Collection (GCC), the GNU Binary Utilities (Binutils), and the GNU command-line utilities (Coreutils, grep, sed, Findutils, etc.)—all this written in C and C++. How does one build the first GCC though? Historically, distributions such as Debian would informally rely on previously-built binaries to build the new ones: when GCC is upgraded, it is built using GCC as available in the previous version of the distribution.

The functional build model does not allow us to "cheat": the whole dependency graph has to be described and be self-contained. Thus, it must describe how the first GCC and C library are obtained. Initially, Guix would rely on of pre-built statically-linked binaries of GCC, Binutils, libc, and the other packages mentioned above to get started [10]. Even though these *binary seeds* were eventually built with Guix and thus reproducible and verifiable using the same Guix revision, they were just that: around 250 MiB of opaque, non-auditable binaries.

In 2017, Nieuwenhuizen *et al.* sought to address this forty-year-old problem at its root: by ensuring no opaque binaries appear at the bottom of the package dependency graph—no less [9, 32]. To that end, Nieuwenhuizen developed GNU Mes, a small interpreter of the Scheme language written in C, capable enough to run MesCC, a non-optimizing C compiler. MesCC is then used to build TinyCC, a more sophisticated C compiler written in C, in turn used to build an old version of GCC, until we get to the modern GCC, written in C++. That, coupled with other efforts, led to a drastic reduction of the size of the opaque binaries at the root of the Guix package graph, well below what had been achieved so far [9, 31]. While many considered it unrealistic a few years earlier, the initial goal of building *everything* from source, starting from a small program of a few hundred bytes and incrementally building more complex pieces of software became a reality in May 2022 [33, 34]. This has the potential to thwart an entire class of software supply chain attacks that has been known but left unaddressed for forty years.

Bootstrapping issues like these do not exist solely at the level of the C language; they show up in many compilers and occasionally in build systems too [47]. Several of them were addressed in Guix for the first time: the Java development kit (JDK) is entirely built from source [46], and so are the Rust [30] and OCaml compilers [9].

## 3 Rationale

As we have seen, Guix is conceptually a source-based distribution. It addresses common classes of software supply chain attacks in two ways: by reducing and eventually removing reliance on opaque binaries at the root of its dependency graph, and by affording reproducible builds. Guix users can choose to obtain pre-built binaries for software they install, and reproducible builds guarantee that anyone can verify that providers of those binaries are not distributing modified or malicious versions.

The security issue that the remainder of this paper focuses on is that of *distributing updates securely*: how can users know that updates to Guix and its package collection that they fetch are genuine? The problem of securing software updates is often viewed through the lens of binary distributions such as Debian, where the main asset to be protected are binaries themselves [3]. Guix being a source-based distribution, the question has to be approached from a different angle.

Guix consists of source code for the tools as well as package definitions that make up the GNU/Linux distribution. Package definitions contain the URL and cryptographic hash of their source code; it is up to package developers writing those definitions to authenticate upstream's source code, for instance by verifying OpenPGP signatures.

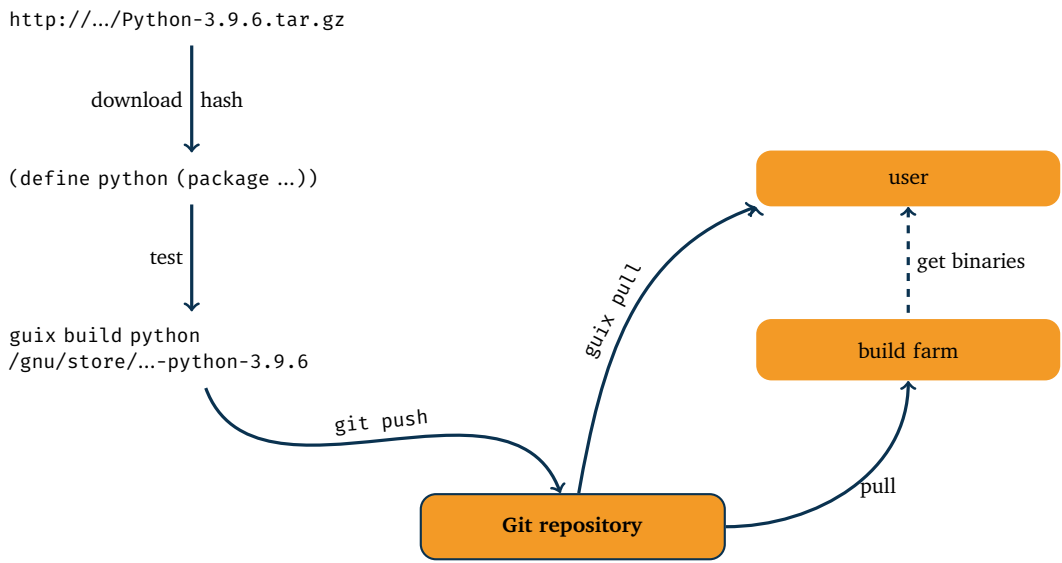

■ **Figure 2**  Supplying software with Guix: developers (left) write package definitions that contain a cryptographic hash of the source code, test them, and publish them in the Git repository; users (right) update their copy from Git using `guix pull` and either fetch binaries for the packages they need or build them locally.

All this code is maintained under version control in a Git repository. To update Guix and its package collection, users run `guix pull`—the equivalent of `apt update` in Debian. When users run `guix pull`, what happens behind the scene is equivalent to `git clone` or `git pull`. This workflow is illustrated in Figure 2.

There are several ways this update process can lead users to run malicious code. An attacker could trick the user into connecting to an alternate repository that contains malicious code or definitions for backdoored packages. This is made more difficult by the fact that code is fetched over HTTPS by default, which allows clients to authenticate the server they are connecting to. However, server authentication is of no use when the server hosting the repository is compromised, as happened to GNU's Savannah in 2010 [19].

An attacker who gained access to the server hosting the Guix repository can push code, which every user would then pull. The change might even go unnoticed and remain in the repository forever. This cannot be addressed simply by having users check OpenPGP signatures on commits: that alone is of no use if users cannot tell whether the commit is indeed signed by an authorized party. An attacker may also reset the main branch to an earlier revision, leading users to install outdated software with known vulnerabilities—a *downgrade attack* [3, 25, 44]. Likewise, the attacker may change the main branch reference so it points to a different branch, containing new malicious code—a *teleport attack* [44].

To summarize, we want to protect against attacks that could be made by gaining access to the Git repository of Guix: introduction of malicious changes by the attacker, downgrade attacks, and teleport attacks. We do *not* aim to protect against cases where an attacker is able to impersonate an authorized developer or otherwise force them into pushing malicious changes; in our model, authorized developers are ultimately trusted.

## 4  Authenticating Git Checkouts

Taking a step back, the problem we are trying to solve is not specific to Guix and to software deployment tools: it's about *authenticating Git checkouts*. By that, we mean that when `guix pull` obtains code from Git, it should be able to tell that all the commits it fetched were pushed by authorized developers of the project. We are really looking at individual commits, not tags, because users can choose to pull at arbitrary points in the commit history of Guix and of third-party channels. Surprisingly, we found that similar Git-backed source-based deployment tools such as Nix do not address this problem, and there were no existing tools or protocols supporting off-line checkout authentication to our knowledge—we get back to that in Section 9.

Git is an append-only, content-addressed version control system. Revision history in Git is represented by a graph of commit objects: each commit can have zero or more parent commits. In the common case, there is a single parent commit pointing to the previous revision; "merge commits", which are created by merging the history of two development branches, have two parents. "Append-only" means that one only ever *adds* new commits to the graph. "Content-addressed" corresponds to the fact that commits are referred to by their cryptographic content hash, currently computed with SHA-1 (more on that in Section 8.2); the contents of a revision, *trees* in Git parlance, as well as other kinds of data stored in a Git repository, are all content-addressed. As an exception, metadata such as references to the latest commit of a branch, is *not* stored in the content-addressed store [44].

Git supports *signed commits*. A signed commit includes an additional header containing an ASCII-armored OpenPGP signature computed over the other headers of the commit. By signing a commit, a Guix developer asserts that they are the one who made the commit; they may be its author, or they may be the person who applied somebody else's changes after review. Checkout authentication requires cryptographically signed commits. It also requires a notion of authorization: we do not simply want commits to have a valid signature, we want them to be signed by an authorized key. The set of authorized keys changes over time as people join and leave the project. The authentication mechanism must be able to deal with those changes; a developer's signature may only be considered valid for the period during which the developer was officially an authorized committer.

The model we devised for *checkout authentication* can be described as *in-band commit authorization*. "In-band" means that the information necessary to determine whether a commit is legitimate—where it was *authorized*—is available in the repository itself. This check can thus be made off-line, without resorting to a third party; it can still be made on a copy of the repository, including an archived copy, years later. Authorization information follows the commit graph: the list of authorized signers for a commit is obtained *in the parent commit(s)*.

To implement this model, we came up with the following mechanism and rule:

1. The repository contains a `.guix-authorizations` file that lists the OpenPGP key fingerprints of authorized committers.
2. A commit is considered authentic if and only if it is signed by one of the keys listed in the `.guix-authorizations` file of each of its parents. We call this the *authorization invariant*.

The `.guix-authorizations` format is a Lisp-style s-expression, as shown in Figure 3. Such a structured format leaves room for extensions, such as per-file authorizations.

```
(authorizations
  (version 0)                    ;current file format version

  (("AD17 A21E F8AE D8F1 CC02  DBD9 F8AE D8F1 765C 61E3"
    (name "alice"))
   ("2A39 3FFF 68F4 EF7A 3D29  12AF 68F4 EF7A 22FB B2D5"
    (name "bob"))
   ("CABB A931 C0FF EEC6 900D  0CFB 090B 1199 3D9A EBB5"
    (name "charlie"))))
```

■ **Figure 3**  Example `.guix-authorizations` file listing authorized committers.

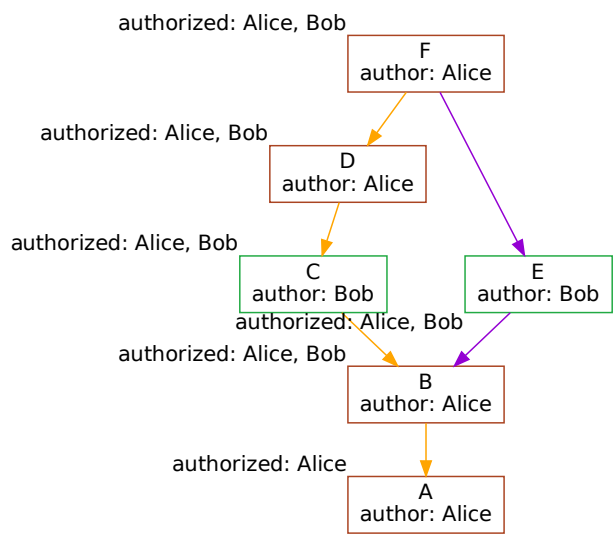

■ **Figure 4**  Graph of commits and the associated authorizations (commit *A* is the parent of commit *B*, commit *B* is the parent of *C* and *E*, and so on).

Let us take an example to illustrate the authorization invariant. In Figure 4, each box is a commit, and each arrow is a parent relationship. This figure shows two lines of development: the orange line on the left may be the main development branch, while the purple line may correspond to a feature branch that was eventually merged in commit *F*. *F* is a merge commit, so it has two parents: *D* and *E*.

Labels next to boxes show who is in `.guix-authorizations`: for commit A, only Alice is an authorized committer, and for all the other commits, both Bob and Alice are authorized committers. For each commit, the authorization invariant holds; for example:

- commit $B$ was made by Alice, who was the only authorized committer in its parent, commit $A$;
- commit $C$ was made by Bob, who was among the authorized committers as of commit $B$;
- commit $F$ was made by Alice, who was among the authorized committers of both parents, commits $D$ and $E$.

The authorization invariant has the nice property that it is simple to state, and simple to check and enforce. This is what `guix pull` implements. If a user's current Guix revision (as returned by the `guix describe` command) is at commit $A$ and the user wants to pull to commit $F$, `guix pull` traverses all these commits and checks the authorization invariant.

Once a commit has been authenticated, all the commits in its transitive closure are known to be already authenticated. `guix pull` keeps a local cache of the commits it has previously authenticated, which allows it to traverse only new commits. For instance, if you are at commit $F$ and later update to a descendant of $F$, authentication starts at $F$.

Since `.guix-authorizations` is a regular file under version control, granting or revoking commit authorization does not require special support. In the example above, commit $B$ is an authorized commit by Alice that adds Bob's key to `.guix-authorizations`. Revocation is similar: any authorized committer can remove entries from `.guix-authorizations`. Key rotation can be handled similarly: a committer can remove their former key and add their new key in a single commit, signed by the former key. If a developer's key is compromised, for instance because their laptop was stolen, they can notify other committers so they quickly remove the key, thereby preventing it from being used to push new commits; the Git log provides an undeniable audit trail when looking for commits that might have been pushed by an attacker who gained access to the key.

The authorization invariant satisfies the needs of the Guix project. It has one downside: it does not play well with the pull-request-style workflow popularized by some source code hosting platforms. Indeed, merging the branch of a contributor not listed in `.guix-authorizations` would break the authorization invariant, unless the committer who accepts the changes signs them, which involves rewriting the commit history of the branch that was submitted. Doing this is possible but it requires the committer to perform this operation on their machine, as opposed to delegating it to the Web server by clicking on the "merge" button. It is a good tradeoff for Guix where the contribution workflow relies on patches sent by email to a patch tracker, and where commits are signed by the committer rather than the original author, but it may be less suitable for other workflows.

## 5 Establishing Trust

You may have noticed that something is missing from the explanation above: what do we do about commit *A* in Figure 4? In other words, which commit do we pick as the first one where we can start verifying the authorization invariant?

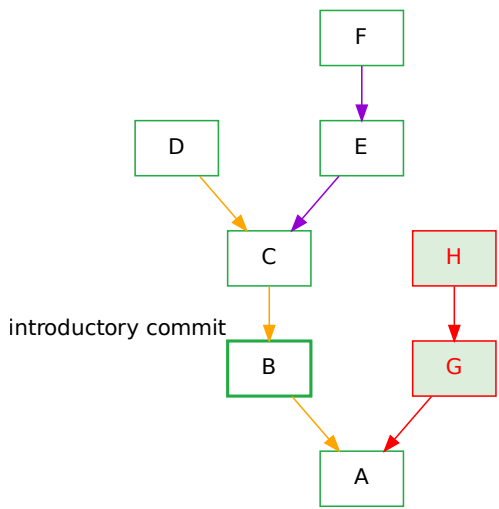

■ **Figure 5** The introductory commit in a commit graph.

We solve this bootstrapping issue by defining *channel introductions*. Previously, one would identify a channel solely by its URL. Now, when introducing a channel to users, one needs to provide an additional piece of information: the first commit where the authorization invariant holds, and the fingerprint of the OpenPGP key used to sign that commit (the fingerprint is not strictly necessary from a security perspective but it provides an additional check).

Consider the commit graph on Figure 5. On this figure, *B* is the *introductory commit*. Its ancestors, such as *A*, are considered authentic. To authenticate *C*, *D*, *E*, and *F*, we check the authorization invariant. Commits *G* and *H* are considered inauthentic because they are not descendants of the introductory commit, *B*.

As always when it comes to establishing trust, distributing channel introductions is very sensitive. The introduction of the official `guix` channel is built into Guix. Users obtain it when they install Guix the first time. Installation instructions tell users to verify the provided OpenPGP detached signature on the tarball or ISO installation image they download. This reduces the chances of getting the "wrong" Guix, following a trust-on-first-use (TOFU) approach.

Guix supports third-party channels providing extra software packages. To use a third-party channel, one needs to add it to the `~/.config/guix/channels.scm` configuration file, which contains a declarative Scheme code snippet listing the desired channels. Authors of third-party channels can also benefit from the channel authentication mechanism: they need to sign commits, to include a `.guix-auth-`

```
(channel
  (name 'my-channel)
  (url "https://example.org/my-channel.git")
  (introduction
    (make-channel-introduction
      "6f0d8cc0d88abb59c324b2990bfee2876016bb86"
      (openpgp-fingerprint
        "CABB A931 C0FF EEC6 900D  0CFB 090B 1199 3D9A EBB5"))))
```

■ **Figure 6**  Specification of a channel along with its *introduction*.

orizations file and the list of relevant OpenPGP keys, and to advertise the channel's introduction. Users then have to provide the channel's introduction in their channels.-scm file, as shown in Figure 6.

The `guix describe` command prints the introduction if there's one. That way, one can share their channel configuration, including introductions, without having to be an expert.

Channel introductions also solve another problem: *forks*. Forks are an integral part of free software, which gives everyone the right to distribute modified copies of the software; one might choose to distribute a fork of Guix or a fork of a channel with different features or different packages. Respecting the authorization invariant "forever" would effectively prevent "unauthorized" forks—forks made by someone who is not in .guix-authorizations. To address this, someone publishing a fork advertises a new introduction for their fork, pointing to a different starting commit.

## 6  Downgrade Attacks

An important threat for software deployment tools is *downgrade attacks,* also called *roll-back* or *replay* attacks [3, 25]. The attack consists in tricking users into installing older, known-vulnerable software packages, which in turn may offer new ways to break into their system. This is not strictly related to the authentication issue discussed above, but it is an important issue that is more easily addressed with this model in place.

Guix saves information about its own provenance—the Git commits of the channels used by `guix pull`. The `guix describe` command prints that information:

```
$ guix describe
Generation 201  Jan 12 2022 18:15:13    (current)
  guix 0052c3b
    repository URL: https://git.savannah.gnu.org/git/guix.git
    branch: master
    commit: 0052c3b0458fba32920a1cfb48b8311429f0d6b5
```

In other words, the `guix` command being used was built from commit `0052c3b...` of the official Git repository. Once `guix pull` has retrieved the latest commit of the selected branch, it can thus verify that it is doing a *fast-forward update*, in Git parlance—just like `git pull` does, but compared to the previously-deployed Guix. A fast-forward update is when the new commit is a descendant of the current commit. Going back to the figure above, going from commit *A* to commit *F* is a fast-forward update, but going from *F* to *A* or from *D* to *E* is not.

Doing a non-fast-forward update would mean that the user is deploying an older version of the Guix currently used, or deploying an unrelated version from another branch. In both cases, the user is at risk of ending up installing older, vulnerable packages. By default `guix pull` errors out on non-fast-forward updates, thereby protecting from roll-backs. Users who understand the risks can override that by passing `--allow-downgrades`.

This does not protect against all forms of *branch teleport attacks* as described by Torres-Arias *et al.* [44]. Specifically, an attacker with access to the server hosting the Git repository could modify the reference of the `master` branch so that it points to an existing development branch that derives from `master`. Users running `guix pull` would upgrade to that branch without problems—it is a fast-forward update. Development branches are usually infrequently merged with `master` and do not receive package security updates very often; consequently this attack could lead users to install outdated packages [16]. Users may not notice the attack because, as long as the branch is active, `guix pull` would still retrieve new changes. However, it would be difficult to hide from developers, which makes the attack less attractive.

Downgrade prevention has been extended to system deployment. When deploying a system with `guix system reconfigure` or a fleet or systems with `guix deploy`, the currently used channels are recorded in the deployed system, as can be seen by running `guix system describe`:

```
$ guix system describe
Generation 161 Apr 27 2021 22:04:13 (current)
  file name: /var/guix/profiles/system-161-link
  canonical file name: /gnu/store/dyx1j...-system
  label: GNU with Linux-Libre 5.11.16
  bootloader: grub-efi
  root device: label: "root"
  kernel: /gnu/store/k029d...-linux-libre-5.11.16/bzImage
  channels:
    guix:
      repository URL: https://git.savannah.gnu.org/git/guix.git
      branch: master
      commit: d904abe0768293b2322dbf355b6e41d94e769d78
  configuration file: /gnu/store/m8rql...-configuration.scm
```

This is useful information for users who wish to map a deployed system back to its source code. We take advantage of that information to implement *system downgrade prevention*: like `guix pull`, deploying a system with `guix system reconfigure` or `guix deploy` now fails with an error when attempting a non-fast-forward update. To our knowledge, this is the first time downgrade prevention is implemented at this level.

## 7 Mirrors and the Risk of Staleness

For package managers, mirrors of the official repositories are a known security risk [3]. Authentication and roll-back prevention as described above allow users to safely refer to mirrors of the official Git repository of Guix. If the official repository at `git.savannah.gnu.org` is down, one can still update by fetching from a mirror, for instance with:

```
guix pull --url=https://github.com/guix-mirror/guix
```

If the repository at this URL is behind what the user already deployed, or if it's not a genuine mirror, `guix pull` will abort. In other cases, it will proceed.

Unfortunately, there is no way to answer the general question "*is X the latest commit of branch B?*". Rollback detection prevents just that, rollbacks, but there is no mechanism in place to tell whether a given mirror is stale. To mitigate that, channel authors can specify, in the repository, the channel's *primary URL*. This piece of information lives in the `.guix-channel` file, in the channel's repository, so it's authenticated. `guix pull` uses it to print a warning when the user pulls from a mirror:

```
$ guix pull --url=https://github.com/guix-mirror/guix
Updating channel 'guix' from 'https://github.com/guix-mirror/guix'...
Authenticating channel 'guix', commits 9edb3f6 to 3e51f9e...
guix pull: warning: pulled channel 'guix' from a mirror of
  https://git.savannah.gnu.org/git/guix.git, which might be stale

Building from this channel:
  guix      https://github.com/guix-mirror/guix 3e51f9e
...
```

Together with downgrade prevention, it makes it more difficult to trick users into getting stale revisions.

## 8 Implementation

Channel authentication as described above is now used in production. This section documents the reasoning behind some of the important implementation choices that we made and our early experience; a companion software artifact is available [12].

### 8.1 Commit Signatures

We chose to use detached OpenPGP signatures on commits. This choice was not motivated by a belief that OpenPGP is the "right tool for the job"—on the contrary, its complexity, which is a result its broad and extensible specification [2], made it a poor candidate in our eyes. More focused options such as minisign [13] looked more appealing. However, we felt that the fact that OpenPGP commit signing is well-supported by Git[2] makes a significant practical difference: developers can easily be set up to sign commits with GnuPG and commands such as `git log` can verify and display signatures; ways to deal with OpenPGP keys and signatures, although complex, are also well-documented.

Key distribution is an important issue. We did not want the whole mechanism to lazily fetch public keys from key servers: this was bound to be unreliable and slow. We instead chose to store keys inside the repository, as plain binary or ASCII-armored OpenPGP packets. Our recommendation is to keep them in a dedicated branch to avoid cluttering regular branches (channel authors can include in the channel metadata the name of the branch where keys are to be found). The authentication code loads keys in memory when it starts and looks them up for signature verification. All the keys ever used to sign commits must be kept in the repository so that past commits can be authenticated. Guix today has 81 ASCII-armored keys representing less than 2 MiB. If needed, this could be reduced by removing unused OpenPGP packets from the keys, such as signature packets, and by storing them in binary format.

## 8.2 Notes on SHA-1

We cannot really discuss Git commit signing without mentioning SHA-1. The venerable crytographic hash function is approaching end of life, as evidenced by recent breakthroughs [27, 42]. Signing a Git commit boils down to signing a SHA-1 hash, because all objects in the Git store are identified by their SHA-1 hash.

Git now relies on a collision attack detection library to mitigate practical attacks [41]. Furthermore, the Git project is planning a hash function transition to address the problem [36].

Some projects such as Bitcoin Core choose to not rely on SHA-1 at all. Instead, for the commits they sign, they include in the commit log the SHA512 hash of the tree, which the verification scripts check [37]. Computing a tree hash *for each commit* in Guix would probably be prohibitively costly. It also would not not address the fact that *every* Git object, not just trees but also commit objects and "blobs" (file contents), is SHA-1-addressed. For now, for lack of a better solution, we rely on Git's collision attack detection and look forward to Git's transition to a more robust hash function.

As for SHA-1 in an OpenPGP context [2]: our authentication code rejects SHA-1 OpenPGP signatures, as recommended [27].

## 8.3 Performance

The core idea, the authorization invariant, is simple to understand and its implementation can be relatively simple, too—a good property for security-sensitive code.

---

[2] When we started this work, the command-line Git tool would only support OpenPGP signatures. Since the release of Git 2.34.0 in November 2021, one can additionally sign commits and tags with OpenSSH, the secure shell client [6].

However, with more than a thousand commits pushed to Guix every month, users may often find themselves authenticating hundreds of commits when running `guix pull`. The implementation must be able to perform well.

At the algorithmic level, the main optimization is to consider that, if a commit has been authenticated, then all the commits in its transitive closure are automatically considered authentic and do not need to be checked. This optimization stems from the fact that the commit graph has integrity properties similar to that of a Merkle-style directed acyclic graph [29]. The implementation takes advantage of it in two ways: by skipping commits that are in the transitive closure of the currently-used Guix commit, and by maintaining a per-user cache of previously-authenticated commits that can also be skipped.

At the implementation level, two key decisions were made: verifying signatures in-process, and dismissing unnecessary OpenPGP features. The go-to technique of spawning GnuPG and Git processes to verify each commit signature would have been prohibitively expensive. Instead, to traverse the Git commit graph, we use libgit2, a C library that implements the Git "protocols" *via* its Guile-Git bindings.

We also have an OpenPGP implementation for GNU Guile, the implementation language of Guix. This OpenPGP implementation is limited to parsing the OpenPGP packets found in signatures and in keys, and to verifying signatures. It does not implement the more complex OpenPGP features that are useless in this context, such as: key signatures and the web of trust, and key expiration and revocation. Timestamps in OpenPGP signatures and expiry dates are easily forged; what matters in our context is the causality of commits: that a signature on a commit is valid and authorized. Likewise, revocation makes little sense in this context; what matters is whether the authorization invariant holds.

On a recent x86_64 laptop (Intel i7 CPU at 2.6 GHz with data stored on a solid state device, SSD), our code authenticates between 600 and 700 commits per second. There are currently between 1,000 and 2,000 commits per month on average (by comparison, the Nixpkgs distribution peaks at about 4,000 commits per month and the Linux kernel, one of the most active free software projects, reaches about 6,000 commits per month). Someone running `guix pull` once per month experiences a 2–3 second delay due to authentication. This does not appear to be detrimental to the user experience.

Another performance aspect has to do with Git repository handling. The mechanisms we devised for commit authentication and downgrade prevention assume the availability of a local copy of the Git repository, including its history. The first time a user invokes `guix pull`, the command clones the repository, downloading more than 300 MiB—this can take minutes, much longer than the commit authentication phase. Currently that operation performs a full clone, including the whole repository history, but it may be possible to optimize it by performing a *shallow clone*, where only recent history is retrieved. Subsequent runs are much faster and lightweight, as Git is able to download just what is missing from the local copy.

## 8.4 Generalization

As explained in Section 3, the general problem being solved — authenticating Git checkouts—is in no way specific to Guix, and the solution we devised may in fact be of interest to *any* Git user. For this reason, and also to facilitate the work of Guix channel developers, we introduced a new command, separate from the channel machinery, to authenticate a Git checkout. The command can be invoked on any Git repository, as in this example:

```
guix git authenticate \
  0c119db2ea86a389769f4d2b9c6f5c41c027e336 \
  "3CE4 6455 8A84 FDC6 9DB4  0CFB 090B 1199 3D9A EBB5"
```

The command above authenticates the checkout in the current directory. The arguments represent its *introduction*: the introductory commit, and the fingerprint of the OpenPGP key used to sign that commit. Additional options allow users to specify, for instance, the name of the branch where OpenPGP keys are to be found.

This command can also authenticate *historical commits*—signed commits made *before* a `.guix-authorizations` file was introduced in the repository. In that mode, users must provide an authorization file that represents the static set of authorizations for all those commits whose parent(s) lack `.guix-authorizations`. We found it useful to retroactively authenticate the history of the Guix repository, where commit signing became compulsory several years before this authentication mechanism was in place.

This interface is low-level and would benefit from simplifications. For instance, repository introductions obtained by users could be gathered in a single place, once for all, such that users do not have to specify them every time. Communicating introductions could also be simplified: the two twenty-byte strings above could be represented as a single 56-character base64 string, or as a QR code. For broad adoption, the best option would be to integrate the functionality in Git proper.

## 8.5 Evaluation

Channel authentication as described above has been deployed and used in production in Guix for more than a year, since June 2020. This has given us an informal but large-scale, "real-world" evaluation of this work. It corresponds, today, to more than 32,000 commits that respect the authorization invariant. The set of authorized developer keys in Guix changed a dozen times in that time frame. The public discussion and review process among developers for the design and implementation of this mechanism [11] helped improve it, eventually leading to a proposal to address a limitation of downgrade prevention [16] and a minor bug fix.

When authentication support was integrated in production code, users who ran `guix pull` transparently obtained the new code, and all subsequent invocations of `guix pull` performed code authentication and downgrade prevention. In almost two years, there was one incident where a committer mistakenly pushed a commit signed with an unauthorized key, which was immediately detected by anyone who attempted to run `guix pull`; the offending commit was removed in minutes (with a *hard reset* to its parent commit, in Git parlance). Such mistakes can be avoided by having a server-side hook running `guix git authenticate`, but we did not have the ability to run such hooks at the time.

Downgrade prevention has had a more visible impact, at least for advanced users with rather unusual workflows. As an example, we have had reports of users who would pull to development or work-in-process branches, using `guix pull --branch=devel`, where `devel` is the name of the branch. When trying to pull back to the main branch, `guix pull` would report an error saying that the target commit is "unrelated" to the source commit. Indeed, because the development branch has not been merged into the main branch, the latest commit on the main branch is not a descendant of the latest commit on the development branch. Since this mechanism is in production, we had only two reports from advanced users "surprised" that switching branches in such a way would trigger the downgrade prevention mechanism; these users were familiar with Git and understood that the mechanism rightfully protected them from a potential downgrade.

System downgrade prevention has demonstrated its value. Since `guix system reconfigure` and `guix deploy` prevent downgrades, a system administrator cannot mistakenly reconfigure the system to an older or unrelated commit; this is particularly useful on systems administered by several people, where an administrator cannot "undo" the upgrade performed by another administrator.

More importantly, checkout authentication together with system downgrade prevention enabled us to provide a trustworthy *unattended upgrade* service. This functionality is typically depended on by server administrators. The service periodically pulls and reconfigures the system. With the guarantees Guix provides, the worst that can happen is that an upgrade does not take place.

Since it became available, authors of Guix channels quickly adopted authentication support. These people were typically already familiar with OpenPGP and signed commits; understanding the authorization model and coming up with a `.guix-authorizations` file was not a barrier to them. Outside Guix, generalized authentication support offered by `guix git authenticate` has seen use in a few repositories. We have yet to see broader adoption but we reckon that simplifying the interface may be a precondition, as explained above. At a more fundamental level, *explaining* that no, signed commits and "verified" tags on a Web user interface (see below) are not enough to authenticate Git checkouts is arguably a prerequisite before we can advocate for a solution.

## 9 Related Work

A lot of work has gone into securing the software supply chain, often in the context of binary distributions, sometimes in a more general context; recent work also looks into Git authentication and related issues. This section attempts to summarize how Guix relates to similar work that we are aware of in these two areas.

**Package manager updates.** The Update Framework [39] (TUF) is a reference for secure update systems, with a well-structured specification [4] and a number of implementations. Many of its goals are shared by Guix. Among the attacks TUF aims to protect against (Section 1.5.2 of the spec), the downgrade-prevention mechanism described in Section 6 does not, *per se*, address *indefinite freeze attacks* (more on that below).

Mercury is a variant of TUF that intends to protect against downgrade attacks even in the face of compromised repositories [25]. Mercury focuses on package version strings to determine what constitutes a downgrade. This is a restrictive definition of downgrade that relies on presumed conventions used by repository maintainers. In contrast, looking at the package commit graph as described in Section 6 allows us to capture the evolution of packages and of the distribution *as a whole*. However, while our approach requires users to download a complete copy of the Git repository, Mercury has much lower bandwidth requirements.

However, both in its goals and system descriptions, TUF is biased towards systems that distribute binaries as plain files with associated metadata. That creates a fundamental impedance mismatch with the functional deployment model we described in Section 2. As an example, attacks such as *fast-forward attacks* or *mix-and-match attacks* do not apply in the context of Guix; likewise, the *repository* depicted in Section 3 of the spec has little in common with a Git repository. The spec also defines a notion of *role*, but those roles do not match our distribution model. With the authentication model described in Section 4, any authorized committer can touch any file; the model and the `.guix-authorizations` format leave room for per-file authorizations, which could be a way to define fine-grain roles in this context.

**Updates for source-based distributions.** The Nix package manager is "source-based" like Guix and distributes its package definitions as a Git repository. It does not currently implement Git checkout authentication and secure updates. A proposal requiring committers to sign commits was rejected, mainly for two reasons: (1) it would make it impossible to perform Git merges (accepting "pull requests") from the GitHub web interface, and (2) GitHub is effectively considered a trusted third party [5]. Nix also features a newer and more decentralized mechanism to distribute packages called *flakes*; flake authentication has been discussed but those discussions have not come to fruition yet [18].

Other package managers have a similar setup: Brew updates its package repositories from GitHub, using Git, but without any particular mechanism to ensure checkout authenticity [7]; likewise, CONDA-Forge, a set of Git repositories hosting package recipes for the CONDA package manager, does not offer any authentication mechanism [8], and FreeBSD Ports are in a similar situation [38]. The source-based pkgsrc tool, used on NetBSD, updates its set of package recipes using the CVS version control system, which allows neither for authentication nor for integrity checks [15].

Gentoo, a source-based GNU/Linux distribution, stores package definitions in a Git repository where commits are required to be signed by developers; the project maintains a separate list of currently authorized developer keys [14]. Because the list of authorized keys is separate, it is not clear how to verify whether a given commit is signed by a key that was authorized at the time of signature. Another shortcoming is that the recommended method to update one's local copy of the package repository is *not* Git but instead the rsync file synchronization protocol together with OpenPGP signatures of the files made with a special-purpose release key.

Developers of OPAM, the package manager for the OCaml language, adapted TUF for use with their Git-based package repository, later updated to write Conex [28], a separate tool to authenticate OPAM repositories. OPAM like Guix is a source-based distribution and its package repository is a Git repository containing "build recipes". To date, it appears that `opam update` itself does not authenticate repositories though; it is up to users and developers to run Conex.

**Supply chain integrity.** The in-toto framework [45] can be thought of as a generalization of TUF; it aims at ensuring the integrity of complete software supply chains, taking into account the different steps that comprise software supply chains in widespread use such as Debian's. In particular, it focuses on *artifact flow integrity*—that artifacts created by a step cannot be altered before the next step.

Thanks the functional deployment model, Guix has end-to-end control over artifact flow, from source code to binaries delivered to users. Complete provenance tracking gives anyone the ability to *verify* the source-to-binary mapping, or to simply not use the project's official binaries, as discussed in Section 2. Conversely, in-toto's approach to artifact flow integrity assumes a relative disconnect between steps that makes verification hard in the first place. In a sense, in-toto addresses non-verifiability through attestation. SLSA [23] and sigstore [20] take a similar approach, insisting on certification rather than allowing independent verification of each step.

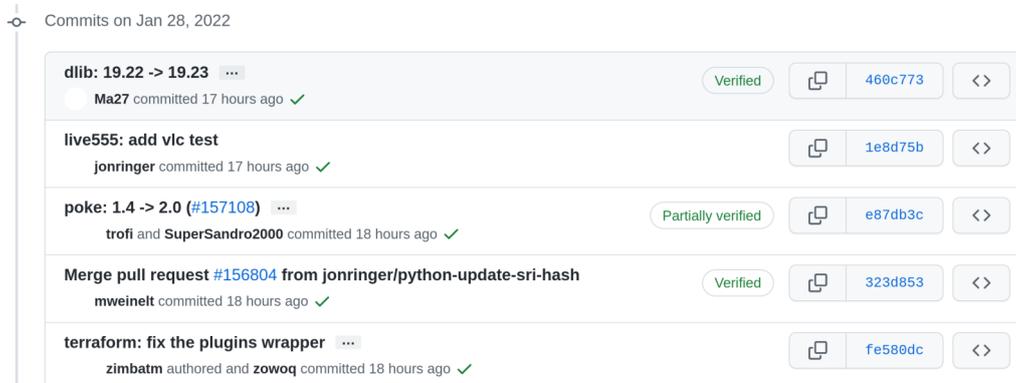

■ **Figure 7** GitHub's Web interface showing commit verification statuses.

**Git authentication.** While signed Git commits (and tags) are becoming more common and generally seen as good practice, we are not aware of other tools or protocols to support off-line Git checkout authentication. Recently, as illustrated in Figure 7, hosting platforms such as GitHub and GitLab started displaying a "verified" tag next to commits signed with the OpenPGP key of the person who pushed them or that of their author—a very limited verification that may give a false sense of security [21, 22]. This mechanism depends on out-of-band data (keys associated with user accounts) and does not permit off-line checks; it also lacks a notion of authorization.

Furthermore, commits made *via* the Web interface are signed by the platform itself, which makes it a single point of trust of every hosted project.

Earlier work focuses on the impact of malicious modifications to Git repository meta-data [44]. An attacker with access to the repository can modify, for instance, branch references, to cause a rollback attack or a "teleport" attack, causing users to pull an older commit or an unrelated commit. As written above, `guix pull` would detect such attacks. However, `guix pull` would fail to detect cases where metadata modification does not yield a rollback or teleport, yet gives users a different view than the intended one—for instance, a user is directed to an authentic but different branch rather than the intended one [16]. This potentially allows for *indefinite freeze attacks*, though these would likely be quickly detected. The "secure push" operation and the associated *reference state log* (RSL) the authors propose would be an improvement.

## 10   Conclusion

The update authentication mechanism described in this article was deployed more than a year ago. Users updating with `guix pull` may have noticed a new progress bar while commits are being authenticated. Apart from that, the change was transparent and our experience so far has been positive. The authentication mechanism is built around the Git commit graph; it is a mechanism to *authenticate Git checkouts* and in that sense is not tied to Guix and its application domain. To our knowledge, this is the first client-side-only Git update authentication mechanism in use.

Guix records the commits of channels used to deploy a set of packages or even a complete operating system. We took advantage of that, together with knowledge of the commit graph of these channels, to prevent downgrade attacks—both when running `guix pull` and when deploying the operating system, which is another distinguishing feature.

The security of the software supply chain as managed by Guix relies on: auditability (every piece of software is built from source), verifiability (the functional model and reproducible builds make it easy to (re)build binaries and check whether they match the source), and secure updates (users updating Guix can only get genuine code vetted by the project). We think this is a solid foundation that addresses common software supply chain issues at their core. This is a radical, novel approach in a time where most related work focuses on certifying each link of the supply chain as opposed to ensuring verifiability.

The security of free operating systems of course also depends on the security of the upstream software packages being distributed. We hope our Git authentication model and/or tool can find its way in upstream development workflows. This would address one of the weakest points in today's development practices.

**Acknowledgements**   The author thanks the anonymous reviewers of subsequent revisions of this paper for insightful suggestions. His gratitude goes to the developers who contributed to the public review process of the authentication mechanism described in this article, and to those who work tirelessly on bootstrapping and reproducibility.

## References

[1]  Joseph Biden. Executive Order on Improving the Nation's Cybersecurity. May 2021. *https://www.whitehouse.gov/briefing-room/presidential-actions/2021/05/12/executive-order-on-improving-the-nations-cybersecurity/*. Last accessed June 2022.

[2]  Jon Callas, Lutz Donnerhacke, Hal Finney, Rodney Thayer. OpenPGP Message Format (RFC 4880). Internet Engineering Task Force (IETF), November 2007. *https://tools.ietf.org/html/rfc4880*.

[3]  Justin Cappos, Justin Samuel, Scott Baker, John H. Hartman. A Look in the Mirror: Attacks on Package Managers. In Proceedings of the 15th ACM Conference on Computer and Communications Security, CCS '08, pp. 565–574, Association for Computing Machinery, 2008. DOI: 10.1145/1455770.1455841

[4]  Justin Cappos, Trishank Karthik Kuppusamy, Joshua Lock, Marina Moore, Lukas Pühringer. The Update Framework Specification. December 2020. *https://github.com/theupdateframework/specification/*. Last accessed June 2022.

[5]  Nix contributors. [RFC 0100] Sign commits. August 2021. *https://github.com/NixOS/rfcs/pull/100*. Last accessed June 2022.

[6]  Git contributors. Git 2.34 Release Notes. November 2021. *https://raw.githubusercontent.com/git/git/master/Documentation/RelNotes/2.34.0.txt*. Last accessed June 2022.

[7]  Brew contributors. Brew source code. January 2022. *https://github.com/Homebrew/brew*. Last accessed June 2022.

[8]  CONDA-Forge contributors. CONDA-Forge Web site. January 2022. *https://conda-forge.org/*. Last accessed June 2022.

[9]  Nathanaëlle Courant, Julien Lepiller, Gabriel Scherer. Debootstrapping Without Archeology: Stacked Implementations in Camlboot. In Programming Journal, 6, February 2022, . DOI: 10.22152/programming-journal.org/2022/6/13

[10]  Ludovic Courtès. Functional Package Management with Guix. In European Lisp Symposium, June 2013. DOI: 10.48550/arXiv.1305.4584

[11]  Ludovic Courtès, GNU Guix contributors. Trustable "guix pull". May 2016. *https://issues.guix.gnu.org/22883*. Last accessed June 2022.

[12]  Ludovic Courtès, GNU Guix contributors. Accepted Artifact for "Building a Secure Software Chain with GNU Guix". Zenodo, June 2022. *https://doi.org/10.5281/zenodo.6581453*

[13]  Frank Denis. Minisign — Simple tool to sign files and verify signatures. 2021. *https://jedisct1.github.io/minisign*. Last accessed June 2022.

[14]  Gentoo developers. Portage Security. January 2022. *https://wiki.gentoo.org/wiki/Portage_Security*. Last accessed June 2022.

[15]  The pkgsrc Developers. The pkgsrc Guide. January 2022. *https://www.netbsd.org/docs/pkgsrc/*. Last accessed June 2022.

[16] Maxime Devos. Getting diverted to non-updated branches: a limitation of the authentication mechanism?. May 2021. *https://issues.guix.gnu.org/48146*. Last accessed June 2022.

[17] Eelco Dolstra, Merijn de Jonge, Eelco Visser. Nix: A Safe and Policy-Free System for Software Deployment. In Proceedings of the 18th Large Installation System Administration Conference (LISA '04), pp. 79–92, USENIX, November 2004.

[18] Eelco Dolstra. Flake authentication. March 2019. *https://github.com/NixOS/nix/issues/2849*. Last accessed June 2022.

[19] Free Software Foundation. Savannah and www.gnu.org Downtime. 2010. *https://www.fsf.org/blogs/sysadmin/savannah-and-www.gnu.org-downtime*. Last accessed June 2022.

[20] The Linux Foundation. sigstore, a new standard for signing, verifying and protecting software. 2021. *https://www.sigstore.dev/*. Last accessed June 2022.

[21] GitHub, Inc.. Managing commit signature verification. 2021. *https://docs.github.com/en/github/authenticating-to-github/managing-commit-signature-verification*. Last accessed June 2022.

[22] GitLab, Inc.. Signing commits with GPG. 2021. *https://docs.gitlab.com/ce/user/project/repository/gpg_signed_commits/*. Last accessed June 2022.

[23] Google, Inc.. Supply-chain Levels for Software Artifacts (SLSA). June 2021. *https://slsa.dev/*. Last accessed June 2022.

[24] Konrad Hinsen. Staged Computation: The Technique You Did Not Know You Were Using. In Computing in Science Engineering, 22(4) , 2020, pp. 99–103. DOI: 10.1109/MCSE.2020.2985508

[25] Trishank Karthik Kuppusamy, Vladimir Diaz, Justin Cappos. Mercury: Bandwidth-Effective Prevention of Rollback Attacks Against Community Repositories. In 2017 USENIX Annual Technical Conference (USENIX ATC 17), pp. 673–688, USENIX Association, July 2017. ISBN: 978-1-931971-38-6

[26] Chris Lamb, Stefano Zacchiroli. Reproducible Builds: Increasing the Integrity of Software Supply Chains. In IEEE Software, 39(2) , March 2022, pp. 62–70. DOI: 10.1109/MS.2021.3073045

[27] Gaëtan Leurent, Thomas Peyrin. SHA-1 is a Shambles: First Chosen-Prefix Collision on SHA-1 and Application to the PGP Web of Trust. In 29th USENIX Security Symposium (USENIX Security 20), pp. 1839–1856, USENIX Association, August 2020. ISBN: 978-1-939133-17-5

[28] Hannes Mehnert, Louis Gesbert. Conex — establishing trust into data repositories. In Proceedings of the ACM OCaml 2016 Workshop, September 2016.

[29] Ralph C. Merkle. Protocols for Public Key Cryptosystems. In Proceedings of the IEEE Symposium on Security and Privacy, pp. 122–134, April 1980. DOI: 10.1109/SP.1980.10006

[30] Danny Milosavljevic. Bootstrapping Rust. December 2018. *https://guix.gnu.org/en/blog/2018/bootstrapping-rust/*. Last accessed June 2022.

[31]   Jan Nieuwenhuizen. Guix Further Reduces Bootstrap Seed to 25%. June 2020. *https://guix.gnu.org/en/blog/2020/guix-further-reduces-bootstrap-seed-to-25/*. Last accessed June 2022.

[32]   Jan Nieuwenhuizen. GNU Mes web site. 2021. *https://gnu.org/software/mes*. Last accessed June 2022.

[33]   Jan Nieuwenhuizen. GNU Mes — the Full Source Bootstrap. February 2021. *https://fosdem.org/2021/schedule/event/gnumes/*. Last accessed June 2022.

[34]   Jan Nieuwenhuizen. Patch series implementing the "full-source bootstrap". May 2022. *https://issues.guix.gnu.org/55227*. Last accessed June 2022.

[35]   Sean Peisert, Bruce Schneier, Hamed Okhravi, Fabio Massacci, Terry Benzel, Carl Landwehr, Mohammad Mannan, Jelena Mirkovic, Atul Prakash, James Bret Michael. Perspectives on the SolarWinds Incident. In IEEE Security & Privacy, 19(02) , Los Alamitos, CA, USA, March 2021, pp. 7-13. DOI: 10.1109/MSEC.2021.3051235

[36]   Git project. Hash Function Transition. 2021. *https://git-scm.com/docs/hash-function-transition/*. Last accessed June 2022.

[37]   BitCoin Core project. Tooling for verification of PGP signed commits. 2021. *https://github.com/bitcoin/bitcoin/tree /d4b3dc5b0a726cc4cc7a8467be43126e78f841cf/contrib/verify-commits*. Last accessed June 2022.

[38]   The FreeBSD Documentation Project. FreeBSD Handbook. January 2022. *https://docs.freebsd.org/en/books/handbook/*. Last accessed June 2022.

[39]   Justin Samuel, Nick Mathewson, Justin Cappos, Roger Dingledine. Survivable Key Compromise in Software Update Systems. In Proceedings of the 17th ACM Conference on Computer and Communications Security, CCS '10, pp. 61–72, Association for Computing Machinery, 2010. DOI: 10.1145/1866307.1866315

[40]   Michael Sperber, R. Kent Dybvig, Matthew Flatt, Anton Van Straaten, Robert Bruce Findler, Jacob Matthews. Revised6 Report on the Algorithmic Language Scheme. In Journal of Functional Programming, 19, 7 2009, pp. 1–301. DOI: 10.1017/S0956796809990074

[41]   Marc Stevens, Daniel Shumow. Speeding up Detection of SHA-1 Collision Attacks Using Unavoidable Attack Conditions. In Proceedings of the 26th USENIX Conference on Security Symposium, SEC'17, pp. 881–897, USENIX Association, 2017. ISBN: 978-1-931-97140-9

[42]   Marc Stevens, Elie Bursztein, Pierre Karpman, Ange Albertini, Yarik Markov. The First Collision for Full SHA-1. In Advances in Cryptology – CRYPTO 2017, pp. 570–596, Springer International Publishing, 2017. DOI: 10.1007/978-3-319-63688-7_19

[43]   Ken Thompson. Reflections on Trusting Trust. In Communications of the ACM, 27(8) , New York, NY, USA, August 1984, pp. 761–763. DOI: 10.1145/358198.358210

[44]   Santiago Torres-Arias, Anil Kumar Ammula, Reza Curtmola, Justin Cappos. On Omitting Commits and Committing Omissions: Preventing Git Metadata

Tampering That (Re)introduces Software Vulnerabilities. In 25th USENIX Security Symposium, pp. 379–395, USENIX Association, August 2016. ISBN: 978-1-931971-32-4

[45] Santiago Torres-Arias, Hammad Afzali, Trishank Karthik Kuppusamy, Reza Curtmola, Justin Cappos. in-toto: Providing farm-to-table guarantees for bits and bytes. In 28th USENIX Security Symposium, pp. 1393–1410, USENIX Association, Aug 2019. ISBN: 978-1-939133-06-9

[46] Ricardo Wurmus. Building the JDK without Java. June 2017. *https://www.freelists.org/post/bootstrappable/Building-the-JDK-without-Java*. Last accessed June 2022.

[47] Ricardo Wurmus, Ludovic Courtès, Paul Wise, Gábor Boskovits, rain1, Matthew Kraai, Julien Lepiller, Jeremiah Orians, Jelle Licht, Jan Nieuwenhuizen. Bootstrappable Builds. 2022. *https://bootstrappable.org/*. Last accessed June 2022.

## About the author

**Ludovic Courtès** works as a research software engineer at Inria, France. He has been contributing to the development of GNU Guix since its inception in 2012 and works on its use in support of reproducible research workflows. You can reach him at *ludovic.courtes@inria.fr*.